\documentclass[9pt,twocolumn,twoside]{opticajnl}
\journal{opticajournal} % use for journal or Optica Open submissions

\setboolean{shortarticle}{true}
\usepackage{layouts}
\usepackage{lineno}
\usepackage{cleveref}
\title{Revisiting the dispersions and quality factors of nanoshell plasmons}

\author[1,*]{Ali Ghorashi}
\author[2]{Victoria Zhang}

\affil[1]{Department of Applied Physics, Yale University, New Haven, Connecticut 06520, USA}
\affil[2]{Department of Physics, Massachusetts Institute of Technology, Cambridge, MA 02139, USA}

\affil[*]{ali.ghorashi@yale.edu}

\begin{abstract}
The dispersions and quality factors of a metallic nanoshell's plasmons may be readily tuned by varying the thickness of the nanoshell. This is because 1. Thickness determines the hybridization between the two nanoshell boundaries and 2. Thickness affects the electronic surface scattering rate and, therefore, the non-radiative decay rates of the plasmonic resonances. We find that prior theoretical works, focused primarily on the thin-nanoshell and retarded regimes, obtained qualitatively incorrect results for both the plasmonic dispersions and the associated plasmonic quality factors. Some of these errors have been pointed out in subsequent works, though the corrected results have not been reported, to our knowledge. Beyond obtaining the correct results (correct dispersions and quality factors as a function of nanoshell thickness), we scrutinize the assumption (used by many prior works) that Matthiessen's rule may be used to combine the effects of bulk and surface-mediated plasmonic scattering processes. To do so, we extend prior work by Dingle to the nanoshell geometry. Although we find qualitative agreement with the Matthiessen's rule approach, we find that Dingle's theory systematically predicts higher quality factors across the range of nanoshell thicknesses we consider. 
\end{abstract}

\setboolean{displaycopyright}{false} % Do not include copyright or licensing information in submission.

\begin{document}
\maketitle
\section{Introduction}
Metallic nanoparticles support purely outgoing solutions to Maxwell's equations, termed localized surface plasmon resonances (LSPRs) \cite{maier2007plasmonics, ruppin1975optical}, at discrete frequencies in the lower half of the complex frequency plane (assuming an $e^{-i\omega t}$ time convention). For particles with spherical symmetry, such as nanospheres and nanoshells, these complex frequencies correspond to poles of the spherical Mie coefficients \cite{bohren2008absorption, jackson1998classical}. The nanoshell Mie coefficients, in particular, may be calculated through recursive application of a transfer matrix, yielding an LSPR dispersion relation for each angular momentum number, $l$ (see Equation 2 of \cite{teperik2004radiative}). In the case of small nanoshells, for which retardation effects may be neglected, an approximate quasistatic dispersion relation may be obtained \cite{alpeggiani2012surface, gao2025high, prodan2003hybridization, maier2007plasmonics}. In our work, however, we focus on the radiative regime (nanoshell radii comparable to the wavelength of light), for which the exact Mie coefficients must be considered. Although the radiative regime has been studied less in the literature, it is relevant for understanding the behavior of lattices of nanovoids \cite{teperik2004radiative, coyle2001confined} and the fundamental limitations of nanoshell plasmons in applications such as biological and environmental sensing \cite{raschke2004gold, sun2002increased}. 

In the first part of this paper, we report the plasmonic dispersions and quality factors of nanoshells in the radiative regime (\cref{Figure: Figure 1} and \cref{Figure: Figure 3}). We find that, for both quantities, previously reported results in the literature are qualitatively incorrect. Crucially, we also find the origins of these errors and reproduce them below (\cref{Figure: Figure 2} and \cref{Figure: Figure 3}). With regards to errors in quality factors, some of our findings were first reported by Moroz \cite{moroz2008electron}, although the corrected results have yet to be reported, to our knowledge.

In the second part of this paper, we scrutinize one of the main assumptions of previous works in determining the quality factors of thin nanoshells: the validity of Matthiessen's rule in combining the electronic bulk and surface scattering rates. We extend the work of Dingle \cite{dingle1950electrical} to nanoshells and find that Dingle's more sophisticated approach qualitatively agrees with Matthiessen's rule but systematically predicts higher quality factors for all considered values of the nanoshell's thickness (\cref{Figure: Figure 4}). 

\section{Plasmonic dispersions and radiative losses for a lossless nanoshell}

\begin{figure}[ht!]
\centering
\fbox{\includegraphics[scale=1]{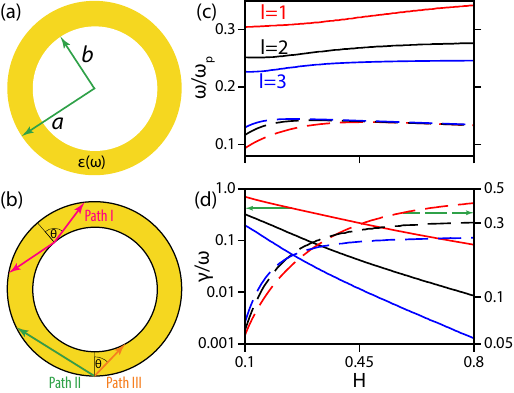}}
\caption{\textbf{Geometry of the scatterer and associated LSPR dispersions/radiative quality factors for $R=6$ and $l=1, 2, 3$: }(a) Nanoshell geometry with inner radius, $b$, outer radius, $a$, and metallic permittivity, $\varepsilon(\omega)$. Both the core and the environment are taken to be vacuum. (b) Relevant surface scattering mechanisms for conduction electrons starting on the inner surface (Path I) and on the outer surface (Path II and Path III). (c) Plasmonic dispersions for each angular momentum channel, $l$. (d) Radiative losses for each angular momentum channel. Dashed (solid) lines correspond to the right (left) vertical axes. Note that (d) shows $\gamma_l/\omega_l$ in accordance with \cite{teperik2004radiative}, which is proportional to the inverse of the plasmonic quality factor.}
\label{Figure: Figure 1}
\end{figure}

We first analyze the case of a nanoshell composed of an ideal lossless metal (in the relevant spectral range) with a Drude permittivity, $\varepsilon(\omega)=1-\omega_p^2/\omega^2$. For ease of comparison, we mostly adopt the notation of \cite{teperik2004radiative}. That is, the inner and outer radii of the nanoshell will be given by $b$ and $a$, respectively (see \cref{Figure: Figure 1}(a)), and, in the following, will appear in the normalized quantities, $H\equiv (a-b)(\omega_p/cl)$ and $R\equiv b\omega_p/cl$, with $c$ being the speed of light in vacuum. In addition, we define the following normalized wavenumbers: $\rho_{a}=a\omega/c$, $\rho_b=b\omega/c$, $\eta_a=\sqrt{\varepsilon}\rho_a$, and $\eta_b=\sqrt{\varepsilon}\rho_b$. In the figures to follow, we report LSPR frequencies, $\omega_l-i\gamma_l$ \footnote{In our first three figures, we show two LSPR branches for each $l$, but we suppress branch indices for brevity.}, as dimensionless quantities, $\omega_l/\omega_p$ and $\gamma_l/\omega_l$. 

To find $\omega_l$ and $\gamma_l$, we use a transfer matrix formalism that is mathematically equivalent to Equation 2 of \cite{teperik2004radiative}. We note, however, that Equation 2 of \cite{teperik2004radiative}, as reported, is not numerically stable for $l>2$ and $|\eta_a|, |\eta_b|\gg 1$. This is due to the exponential behavior of $h_{l}^{(1) ((2))}(\eta_{a, b})$, where $h_{l}^{(1)((2))}$ is the spherical Hankel function of the first (second) kind. We fix this instability by noting that the LSPR dispersion relation depends only on products of Hankel functions (and their derivatives), e.g. terms of the form $h_l^{(1)}(\eta_a)h_l^{(2)}(\eta_b)$. This term in particular may be re-expressed as follows: 
\begin{equation}
h_l^{(1)}(\eta_b)h_l^{(2)}(\eta_a) = \Big[h_l^{(1)}(\eta_b) e^{-i\eta_b}\Big]\Big[h_l^{(2)}(\eta_a) e^{i\eta_a}\Big]\Big[e^{i(\eta_b-\eta_a)}\Big],
\end{equation}
where the first two brackets are non-exponential and the last bracket is a well-behaved exponential because $|\eta_a-\eta_b|\ll |\eta_a|, |\eta_b|$, making the product of the three brackets numerically stable. All poorly behaved terms of the dispersion relation may be tamed in a similar manner.

\begin{figure}[ht!]
\centering
\fbox{\includegraphics[scale=1]{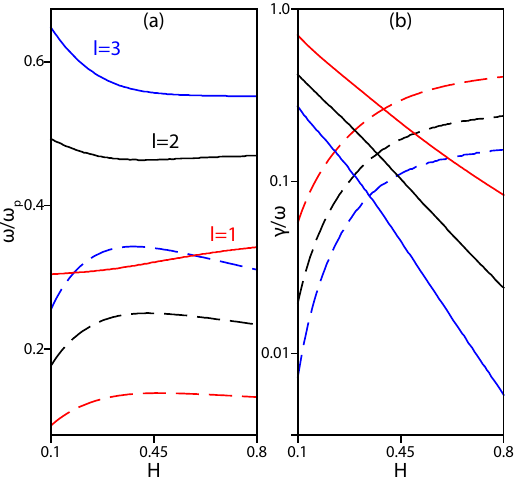}}
\caption{\textbf{Dispersions and radiative losses for $Rl=6$:} (a) Plasmonic dispersions for each angular momentum channel. (b) Radiative losses for each angular momentum channel, $l$. As in \cref{Figure: Figure 1}, no electronic scattering mechanisms are taken into account in this figure, so quality factors are determined entirely by radiative losses. }
\label{Figure: Figure 2}
\end{figure}

We show dispersions and quality factors in \cref{Figure: Figure 1}(c, d) for $R=6$ and $0.1<H<0.8$. In the plotted spectral range, we have two LSPR resonances for each $l$, with the lower (higher) frequency resonances decreasing (increasing) in quality factors as the thickness of the nanoshell is increased. We note that Figure 2 of \cite{teperik2004radiative}, which ostensibly uses the same physical parameters as our \cref{Figure: Figure 1}, disagrees qualitatively with our figure (even in the ordering of some of the modes). We found that their result actually corresponds to $Rl=6$ instead of $R=6$, which is why our $l=1$ results match but our $l>1$ results differ quite strongly \footnote{A clue may be found on page 3 of \cite{teperik2004radiative}, where the authors state "this value of $R$ corresponds to an actual internal radius of $b\approx150 \text{nm}$". $R$ is $l$-dependent, so it is not possible for a fixed value of $R$ to correspond to the same internal radius for different $l$.}. To support this claim, we reproduce their result in \cref{Figure: Figure 2} by taking $Rl=6$, which agrees perfectly with Figure 2 of \cite{teperik2004radiative}. We stress that fixing $Rl=6$ vs $R=6$ is not just an arbitrary choice. For fixed values of $R$ and $H$, the dispersions, for each $l$, correspond to the same ratio of the inner radius to the outer radius, $b/a=R/(R+H)$ (for each $l$). The result of \cite{teperik2004radiative}, however, corresponds to different, $l$-dependent, ratios, $b/a$, for each value of $H$.

\section{Bulk and surface scattering processes}

\begin{figure}[ht!]
\centering
\fbox{\includegraphics[scale=1]{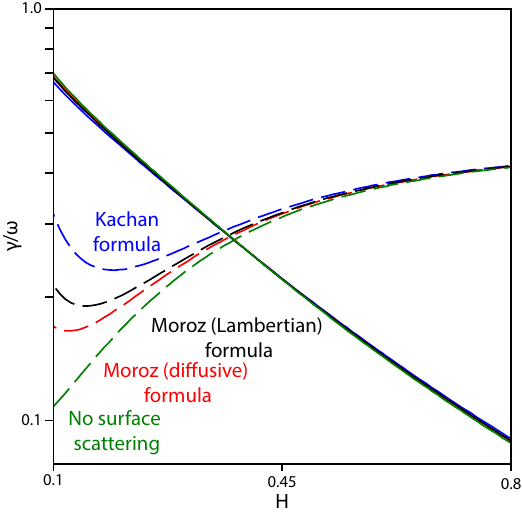}}
\caption{\textbf{Total (radiative + non-radiative) plasmonic losses for $l=1$ and $R=6$:} For the dipolar, $l=1$, channel, we plot the total scattering rate of the plasmonic modes as calculated through Matthiessen's rule with three surface scattering models, as described in the main text. Green line shows plasmonic loss in the absence of surface scattering (but with bulk scattering included).}
\label{Figure: Figure 3}
\end{figure}

We now move on to the general case of a nanoshell composed of a lossy metal. The Drude permittivity in this case will be given by $\varepsilon(\omega)=1-\omega_p^2/(\omega(\omega+i\gamma_e))$, where $\gamma_e$ is the total transport scattering rate of conduction electrons. Conduction electrons may  scatter through bulk processes (involving phonons, other electrons, and impurities) with an associated scattering rate $\gamma_{\text{bulk}}$ \cite{grigorchuk2012radiative}. In addition, they may scatter off of the two nanoshell boundaries, with an associated surface scattering rate $\gamma_{\text{surface}}$. In finding the total scattering rate, $\gamma_e$, most prior works apply Matthiessen's rule \cite{teperik2004radiative, ye2018effect, raschke2004gold, maier2007plasmonics, averitt1999linear, coronado2003surface}$: \gamma_e=\gamma_{\text{bulk}}+\gamma_\text{surface}$. For now, we also use Matthiessen's rule, but we examine the validity of this assumption in the next section. In the following, for concreteness, we take $\gamma_\text{bulk}\approx 0.011\omega_p$ and we calculate $\gamma_\text{surface}$ using diffusive and Lambertian scattering models \cite{moroz2008electron, moroz2011electron}. For both diffusive and Lambertian scattering, we find $\gamma_\text{surface}$ by using the relation $\gamma_\text{surface}=v_F/\lambda$, where $v_F$ is the Fermi velocity (which we take to be $\approx 4.7\times 10^{-3}c$) and $\lambda$ is the average geometric mean free path for a particle starting on one of the nanoshell boundaries, given by the following expression:
\begin{equation}
    \lambda=\frac{b^2}{a^2+b^2}\langle \lambda(\theta, \phi) \rangle_{\Omega_1}+\frac{a^2}{a^2+b^2}\langle \lambda(\theta, \phi) \rangle_{\Omega_2},
    \label{Equation: Geometric mean free path}
\end{equation}
where the first term corresponds to averaging (denoted by $\langle...\rangle$) over all paths originating from the inner boundary, $\Omega_1$ (Path I in \cref{Figure: Figure 1}(b)), and the second term corresponds to averaging over all paths originating from the outer boundary, $\Omega_2$ (Paths II and III in \cref{Figure: Figure 1}(b)). The prefactors in both terms take into account that there are more electrons on the outer boundary than on the inner boundary (with the number of electrons on each surface being proportional to the area of the chosen surface). Note that a given chord length, $\lambda(\theta, \phi)$ is, for the case of a nanoshell, independent of the azimuthal angle, $\phi$, but we keep the dependence to maintain generality.

For diffusive scattering, averaging over the angle $\theta$ is performed with the standard solid angle measure, $\sin(\theta)\text{d}\theta\text{d}\phi$ whereas Lambertian scattering \cite{kreibig1974electronic} has an extra factor of $\cos(\theta)$ in the measure (corresponding to uniform "brightness" of electrons from all lines of sight \cite{moroz2011electron}). 
The first calculation of the diffusive scattering rate was obtained by Kachan \cite{kachan1999limitation, kachan2001resonance}. However, as pointed out by Moroz, their result, which was used in Figure 7 of \cite{teperik2004radiative}, was incorrect. The corrected result by Moroz \cite{moroz2008electron}, which we independently verified both analytically and through numerical integration, gives qualitatively different quality factors as shown in \cref{Figure: Figure 3} below.

In \cref{Figure: Figure 3}, we show the total plasmonic scattering rate, $\gamma_{l}$, for both of the $l=1$ branches shown in \cref{Figure: Figure 1} and \cref{Figure: Figure 2}. \Cref{Figure: Figure 3} shows that the formula by Kachan significantly overestimates the plasmonic decay rate for small values of the dimensionless thickness, $H$. Note that surface scattering is only important when $v_F/(a-b)$ is comparable to $\gamma_\text{bulk}$ \cite{kreibig1969limitation}, i.e., when $(v_F/c)/(H\gamma_\text{bulk}/(\omega_p))\approx 1$. For our chosen Fermi velocities and bulk scattering rates, this means that the surface contribution will be important for approximately $H < 0.4$, which is consistent with \cref{Figure: Figure 3}. 

It is reasonable at this point to wonder which is the "correct" surface scattering model; i.e., the one most consistent with experiments. Experimentalists model $\gamma_{\text{surface}}$ as  $\gamma_{\text{surface}}=Av_F/(a-b)$, with $A$ a dimensionless parameter. $A=0.5$ is consistent with the Lambertian model. However, reported values of $A$ range from $A=0.5$ to $A>1$ \cite{ye2018effect, raschke2004gold, westcott2002relative}. We will return to this question after introducing our last surface scattering model in the next section. 

\section{Transport-limited geometric approach to the mean free path}
In the discussion above, the decay rate of conduction electrons was approximated through Matthiessen's rule, $\gamma_e = \gamma_{\text{bulk}}+\gamma_{\text{surface}}$, where $\gamma_{\text{surface}}$ was obtained by assuming either diffusive or Lambertian scattering. However, such an approximation does not account for the fact that the bulk and surface scattering mechanisms are not, in actuality, independent of one another. By this, we simply mean that each chord length, $\lambda(\theta, \phi)$ in \cref{Equation: Geometric mean free path} should actually be limited by the bulk mean free path of the conduction electrons.

To address this issue, we follow Dingle's approach for thin wires \cite{dingle1950electrical}. That is, we account for the effects of \emph{all} scattering mechanisms by calculating the average \emph{total} mean free path of conduction electrons, $\lambda_{\text{tot}}$, through the following expression:
\begin{equation}
    \frac{\lambda_{\text{tot}}}{\lambda_0}=\Bigg[\frac{b^2\langle  [1-e^{-\lambda(\theta, \phi)/\lambda_0}]\rangle_{\Omega_1}+a^2\langle  [1-e^{-\lambda(\theta, \phi)/\lambda_0}]\rangle_{\Omega_2}}{a^2+b^2}\Bigg],
    \label{Equation: mean free path}
\end{equation}
where $\lambda_0=v_F/\gamma_{\text{bulk}}$ is the mean free path in the absence of surface scattering. After $\lambda_{\text{tot}}$ is calculated, the correct prescription for $\gamma_e$ is clearly $\gamma_e=v_F/\lambda_\text{tot}$. Note that in the $\lambda_0/(a-b)\rightarrow\infty$ limit, \Cref{Equation: mean free path} reduces to 
\begin{equation}
\lambda_{\text{tot}}\Big|_{\lambda_0\gg a-b}\approx \Bigg[\frac{b^2\langle  \lambda(\theta, \phi)\rangle_{\Omega_1}+a^2\langle \lambda(\theta, \phi)\rangle_{\Omega_2}}{a^2+b^2}\Bigg],
    \label{Equation: mean free path 2}
\end{equation}
which is the same as \cref{Equation: Geometric mean free path}, as expected. Of course, in the opposite limit, $\lambda_0/(a-b)\rightarrow 0$, we have $\lambda_{\text{tot}}\rightarrow \lambda_0$. Thus, in the thin shell and thick shell limits, we expect the quality factors obtained using \cref{Equation: mean free path} to coincide with those obtained using \cref{Equation: Geometric mean free path} in conjunction with Matthiessen's rule. 

\begin{figure}[ht!]
\centering
\fbox{\includegraphics[scale=1]{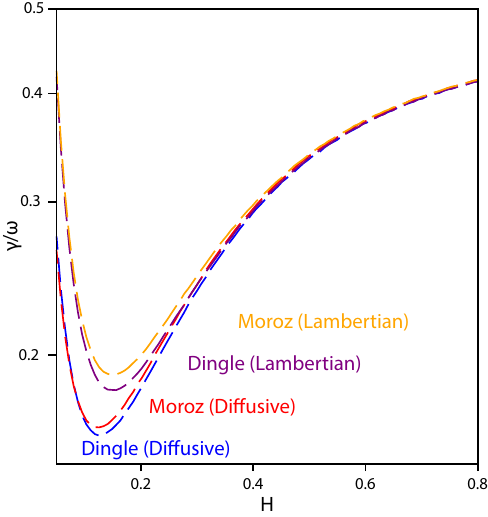}}
\caption{\textbf{Effect of the Dingle correction for $l=1$ and $R=6$:} We plot the total plasmonic losses (for the lower frequency branch of the dipolar modes) as predicted by \cref{Equation: Geometric mean free path} in conjunction with Matthiessen's rule (red and yellow lines) and through Dingle's approximation (see \cref{Equation: mean free path}) (blue and purple lines) for diffusive and Lambertian surface scattering.}
\label{Figure: Figure 4}
\end{figure}

As shown in \cref{Figure: Figure 4}, this is indeed the case. Furthermore, qualitative agreement between the two approaches is also seen in the intermediate thickness regime. We note, however, that Dingle's approach systematically predicts higher quality factors for all nanoshell thicknesses. 

\section{Limitations of our work}

The classical approach employed in our work clearly has the benefit of simplicity. It should be noted, however, that a quantum mechanical approach \cite{kirakosyan2016surface, shahbazyan2018surface, shahbazyan2016landau, kraus1983plasmon} or a full Boltzmann transport theory approach  \cite{sondheimer2001mean, fuchs1938conductivity, kumar2022ultralow, kuznetsova2014effect, semchuk2021kinetic} would be expected to be more accurate, especially for higher angular momentum modes. In particular, the geometric approach used here does not account for the fact that the electric field distribution of the plasmonic mode of interest affects the effective mean free paths of the conduction electrons participating in that plasmonic mode \cite{khurgin2016landau}. There are also systems in which the basic assumptions of the classical approach fail entirely, such as anisotropic nanoshells. Although the resonances of such systems are well understood \cite{wu2009influence, wallen2015anomalous, qiu2010light, yin2011electromagnetic}, the correct approach to analyzing a nanoshell composed of an anisotropic metal (as opposed to an anisotropic core \footnote{Note that \cite{wu2009influence} used a classical surface scattering model for the case of a spherically anistoropic core. This is justified because the metal shell itself is isotropic.}) within the classical geometrical scattering framework has yet to be established, to our knowledge.

We stress that the aim of the present work is to clarify the (correct) predictions made by the different scattering models (diffusive, Lambertian, and with/without Dingle's method) and to quantify how much these predictions differ from one another. The question of which model works in which regime(s) (if any) is still an open question and requires experimental results on high quality nanoshells across a range of nanoshell thicknesses. 

%\author{Author One\authormark{1} and Author Two\authormark{2,*}}

%\address{\authormark{1}Peer Review, Publications Department,
%Optica Publishing Group, 2010 Massachusetts Avenue NW,
%Washington, DC 20036, USA\\
%\authormark{2}Publications Department, Optica Publishing Group,
%2010 Massachusetts Avenue NW, Washington, DC 20036, USA\\
%%\authormark{3}xyz@optica.org}

%\email{\authormark{*}xyz@optica.org}}

%Example with the corresponding author designated by an asterisk and a note indicating equal contributions by two authors.

%\author{Author One\authormark{1,3} and Author %Two\authormark{2,3,*}}

%\address{\authormark{1}Peer Review, Publications Department,
%Optica Publishing Group, 2010 Massachusetts Avenue NW, %Washington, DC 20036, USA\\
%\authormark{2}Publications Department, Optica Publishing Group, %2010 Massachusetts Avenue NW, Washington, DC 20036, USA\\
%\authormark{3}The authors contributed equally to this work.\\
%\authormark{*}xyz@optica.org}}

%\section{Examples of Article Components}
%\label{sec:examples}

%\begin{figure}[ht!]
%\centering\includegraphics{opticafig2}
%\caption{Terahertz focusing metalens.}
%\end{figure}

\section{Back matter}
\begin{backmatter}
\bmsection{Acknowledgment} A.G. thanks Andrea Alu and Ravishankar Sundararaman for useful discussions. V.Z. acknowledges support from the NSF GRFP under Grant Number
2141064.
\smallskip
\bmsection{Disclosures} The authors declare no conflicts of interest.

\bmsection{Data availability} Data underlying the results presented in this paper are publicly available on Github \cite{Coherent-Perfect-Absorption}.

\bigskip

\end{backmatter}

% Bibliography
\bibliography{sample}

% Full bibliography added automatically for Optics Letters submissions; the following line will simply be ignored if submitting to other journals.
% Note that this extra page will not count against page length
\bibliographyfullrefs{sample}

\end{document}